\documentclass[aps,prb,twocolumn,showpacs,amsmath,amssymb]{revtex4}
\usepackage{graphicx}
\usepackage{color}

\usepackage{hyperref}

\begin{document}

\title{Random Fractal Ansatz for the configurations of Two-Dimensional Critical Systems}

\author{Ching Hua Lee${}^{a}$}
\author{Dai Ozaki${}^{b}$}
\author{Hiroaki Matsueda${}^{c}$}
\affiliation{
${}^{a}$Institute of High Performance Computing, 138632, Singapore \\
${}^{b}$Department of Applied Physics, Tohoku University, Sendai, 980-8579, Japan.\\
${}^{c}$Sendai National College of Technology, Sendai 989-3128, Japan
}
\date{\today}
\begin{abstract}
Critical systems have always intrigued physicists and precipitated the development of new techniques. Recently, there has been renewed interest in the information contained in their classical configurations, whose computation do not require full knowledge of the wavefunction. Inspired by holographic duality, we investigated the entanglement properties of the classical configurations (snapshots) of the Potts model by introducing an ansatz ensemble of random fractal images. By virtue of the central limit theorem, our ansatz accurately reproduces the entanglement spectra of actual Potts snapshots without any fine-tuning of parameters or artificial restrictions on ensemble choice. It provides a microscopic interpretation of the results of previous studies, which established a relation between the scaling behavior of snapshot entropy and the critical exponent. More importantly, it elucidates the role of ensemble disorder in restoring conformal invariance, an aspect previously ignored. Away from criticality, the breakdown of scale invariance leads to a renormalization of the parameter $\Sigma$ in the random fractal ansatz, whose variation can be used as an alternative determination of the critical exponent. We conclude by providing a recipe for the explicit construction of fractal unit cells  consistent with a given scaling exponent. 
\end{abstract}
\pacs{05.10.Cc, 05.45.Df, 05.70.Jk, 
07.05.Pj, 11.25.Hf, 11.25.Tq, 75.10.Hk, 89.70.Cf, 89.75.Da}
\maketitle


\section{Introduction}

Critical phenomena are ubiquitous in modern physics, accompanying second-order phase transitions in a wide variety of scenarios. However, they are intrinsically difficult to study numerically due to their divergent correlation lengths. As such, they have spurred the development of techniques like transfer matrix methods and the Renormalization Group approach by Wilson and others. Today, the study of criticality lie at the intersection of important topics like the theory of phase transitions, conformal field theory (CFT), entanglement and holographic duality\cite{maldacena1998,witten1998,gubser1998}.

Given the ubiquity of critical systems, it is desirable to have a numerically inexpensive way of computing their essential properties like their critical exponent and critical temperature. One recently proposed approach employs the snapshot spectrum, which captures the entanglement properties of the classical configurations of the critical system at different scales. Indeed, in Monte Carlo (MC) simulations of strongly correlated spin systems, the classical configurations are much more easily obtained than their wavefunctions. By zooming in onto the physics at different scales, the snapshot spectrum is intrinsically suited for studying scale-invariant phenomena like phase transitions\cite{matsueda2012holographic}. As we shall see, it can also elucidate the role of disorder in restoring the broken conformal symmetry in snapshots of classical configurations. Furthermore, it yields entropic scaling laws reminiscent of those of critical quantum systems, which are intensively studied in holographic duality\cite{trotter1959product,suzuki1976relationship,ryu2006holographic,ryu2006aspects,hartnoll2009,horowitz2009,mcgreevy2010,sachdev2012,huijse2012,sachdev2015bekenstein,witczak2013dynamics,lucas2015memory,bhaseen2013holographic,bhaseen2015energy,baier2008relativistic,cai2015introduction,lucas2014scale,lucas2015hydrodynamic,huang2015entanglement,lee2016exact,gulee2016}. 


In analogy to the entanglement spectrum for quantum systems, the \emph{snapshot spectrum} is the singular-value decomposition (SVD) spectrum of an image (i.e. snapshot) of a classical spin configuration. Its corresponding von-Neumann entropy, known as the \emph{snapshot entropy}, quantifies the complexity of the hierarchical structures within the snapshot. 
In Ref. \onlinecite{matsueda2012holographic,matsueda2015proper}, the onset of criticality in Ising and Potts models were unambiguously identified from the scaling behavior of their snapshot entropies. Subsequently, this scaling behavior was also rigorously shown \cite{imura2014snapshot,matsueda2014comment} to yield the scaling exponent of the system. It remains an open question whether further information, i.e. the central charge of the dual conformal field theory, can be extracted from it. For a related problem involving critical percolation, the answer has turned out to be in the affirmative - certain percolation cluster images do contain information of the central charges for their associated CFTs, as demonstrated through the techniques of Schramm-Loewner evolution (SLE)\cite{kager2004guide,nienhuis2009stochastic,cardy2005sle,saint2009geometric}.

A precise understanding of the information contained in snapshot spectra requires a detailed scrutiny of the role of disorder in snapshots, an aspect not carefully studied in previous works\cite{matsueda2012holographic,imura2014snapshot,matsueda2014comment}. While the full set of conformal symmetries (translation, scaling, rotation and inversion) is present at criticality, each snapshot reflects a ``classical'' configuration in which these symmetries are manifestly broken. To restore these symmetries, one has to take ensemble averages of the snapshots. However, ensemble averaged snapshot spectra (EASS) are inevitably complicated by the ensemble disorder. For instance, the EASS at both critical temperature $T_C$ and very high temperature $T\gg T_C$ look superficially similar to the spectra of random matrices, despite the former having a much higher (conformal) symmetry\cite{imura2014snapshot,matsueda2014comment}. Ensemble disorder has also resulted in marked deviations from the snapshot entropy scaling law $S_\chi \sim a \chi^{\eta}\log \frac{\chi}{b}$ derived in Refs. \onlinecite{imura2014snapshot} and \onlinecite{matsueda2014comment}, where $a$ and $b$ depend on the system size. Hence, the important question to be answered is: What features of the snapshot spectrum and entropy are truly reflective of the conformal symmetry that characterize criticality? 


To help bridge the abovementioned gap in understanding, we introduce an ansatz ensemble of \emph{random fractal} snapshot images that accurately reproduces the EASS of actual systems both at and away from criticality. This agreement is a spectacular testament to the appropriateness of this ansatz, which possess no tunable parameters. With a fixed system size, its only parameter $\Sigma$, which characterizes the microscopic heterogeneity of the ensemble, is fixed by the effective critical exponent of the system. 
As a precisely constructed ansatz, the random fractal ensemble also have the advantage of possessing rigorously known spectral properties which complements numerical results in the literature\cite{matsueda2012holographic,imura2014snapshot,matsueda2014comment,lee2014exact}, such as providing an exact expression for the entropy contribution from each scale. 

This paper is organized as follows. In Section II, we begin with a review of snapshot entropy and spectrum as well as the Ising and Potts models that we simulated. Next, we introduce our random fractal ansatz in Section III, and derive a few general properties. Following that, we specialize in Section IV to ans\"{a}tze possessing identical distributions at different scales, i.e. are scale-invariant. We shall discuss their snapshot spectra and entropy at length, since they are directly relevant to physical critical systems. In Section V, we briefly discuss results away from the critical temperature. Finally in Section VI, we detail the explicit construction of fractal unit cells based on constraints on the scaling exponent.

\section{Preliminaries}
\subsection{The Ising and Potts models}

We first introduce the classical 2-dimensional Q-state Potts model Hamiltonian\cite{blumenhagen2009introduction,francesco2012conformal,matsueda2012holographic}
\begin{equation}
H=-J\sum_{\langle i,j\rangle } \delta_{\sigma_i\sigma_j}
\end{equation}
where each site variable (spin) $\sigma$ takes one of $Q$ possible values. The Hamiltonian assigns an energy penalty of $J$ ($J>0$) for each pair of nearest neighbors sites $i$ and $j$ with dissimilar spins. When $Q=2$, this system is also called the Ising model. 

For $Q\leq 4$, the Potts model undergoes a second-order phase transition between the low temperature (ordered) and high temperature (disordered) phase at the self-dual (critical) temperature of $T_c=2/\log(1+\sqrt{Q})$. This critical regime is of most interest for this paper, with the system governed by a CFT with central charge $c=\frac{2(Q-1)}{Q+2}$. For $Q>4$, $c>1$ and the system undergoes a first-order transition instead.

At criticality, the Potts model also possess correlation functions with well-defined power-law decay exponents $\eta$: $\langle \sigma_i\sigma_j\rangle \sim |r_i-r_j|^{-\eta}$ where $r_{i}, r_j$ are the positions of spins $i$ and $j$. From CFT\cite{francesco2012conformal}, $\eta=2(h+\bar h)$, where $h,\bar h$ are the conformal weights of the holomorphic and anti-holomorphic primary fields $\sigma$ and $\bar \sigma$. These conformal weights are given by
\begin{equation}
h=\bar h=\frac{(pr-p's)^2-(p-p')^2}{4pp'}
\end{equation}
where $(r,s)$ indexes the primary field within the $M(p,p')$ unitary minimal model CFT. For the $Q=2$ (Ising) case, the relevant CFT is $M(4,3)$ with $\sigma$ associated with the $(r,s)=(1,2)$ or $(2,2)$ primary fields with $h=\bar h=\frac1{16}$. This yields the critical exponent of $\eta_{Q=2}=\frac1{4}$. For the $Q=3$ Potts model, the relevant CFT is $M(6,5)$ with $\sigma$ associated with the $(r,s)=(2,3)$ or $(3,3)$ primary fields with $h=\bar h=\frac1{15}$, yielding $\eta_{Q=3}=\frac{4}{15}$. 

We compute classical configurations (snapshots) of the Potts model via Monte Carlo (MC) simulations with up to $10^7$ steps on $L\times L$ square lattices, with $L$ ranging from $128$ to $512$. 


\subsection{Snapshot entropy and spectrum}
\begin{figure}
\includegraphics[scale=0.8]{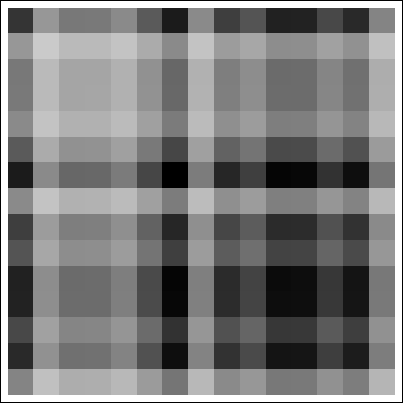}
\includegraphics[scale=0.8]{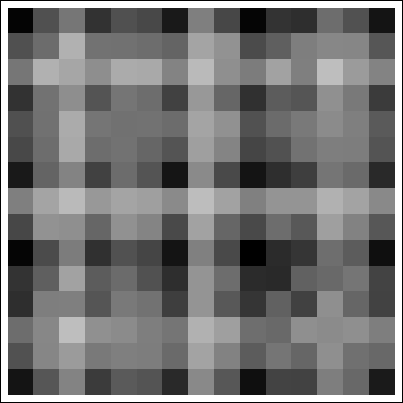}
\caption{(Color online) Left) An example of an image with rank one, i.e. $\Lambda_1=1,\Lambda_2=\Lambda_3=...=0$. It consists of criss-crossing lines and can only contain very limited information. Right) An example of an image with four SVD eigenvalues not close to zero. With a higher rank, it has the possibility of containing more levels of detail. }
\label{SVDexp}
\end{figure}

To characterize the classical configurations (snapshots) of the Potts models, we first review the definitions of the snapshot spectrum and entropy from Ref. \onlinecite{matsueda2012holographic}. Consider an $L_x$ by $L_y$ image $m(x,y)$ of an instantaneous classical configuration of the Potts model. To define the snapshot entanglement spectrum and entropy, we perform partial traces over this image just as in quantum entanglement studies\cite{peschel2002,huang2012,lee2014position}. Partially tracing over the $x$ or $y$ directions, we obtain ``reduced density matrices'' (RDMs)
\begin{equation}
\rho_X(x,x')=\sum_y^{Ly} m(x,y)m(x',y)
\end{equation}
\begin{equation}
\rho_Y(y,y')=\sum_x^{Lx} m(x,y)m(x,y')
\end{equation}
As these RDMs are square matrices, they can each be diagonalized:
\begin{equation}
\rho_X(x,x')=\sum_n^{\text{min}(L_x,L_y)} U_n(x)\Lambda_nU_n(x')
\end{equation}
\begin{equation}
\rho_Y(y,y')=\sum_n^{\text{min}(L_x,L_y)} V_n(y)\Lambda_nV_n(y')
\end{equation}
Their eigenvalues $\Lambda_n$ are equal, as is evident from decomposing the image in terms of the unitary matrices $U_n(x)$ and $V_n(y)$'s:
\begin{equation}
m(x,y)=\sum_n^{\text{min}(L_x,L_y)}U_n(x)\sqrt{\Lambda_n}V_n(y)
\end{equation}
The distribution of $\Lambda_n$ is known as the snapshot spectrum of the image ($\sqrt{\Lambda_n}$ are singular value decomposition (SVD) eigenvalues), with the rank of $m(x,y)$ being the number of nonzero values in $\{\Lambda_n\}$. Images of rank one are the analogs of product states in quantum mechanics (see Fig. \ref{SVDexp}), and are restricted to look like a criss-crossing patchwork. As the rank increases, the image acquires more degrees of freedom and can contain various levels of detail. Snapshots of scale-invariant critical systems contain details of all sizes and orientations, and necessitate a long tail in their SVD spectra, analogous to long-range entanglement in quantum information. In Sect. \ref{regime}, we show that this tail scales like $\Lambda_n\sim n^{\eta-1}$, where $\eta<1$ is the abovementioned decay exponent of \emph{spatial} correlation functions.

\subsection{Representations of spins}
To represent the $Q$ different types of spins in a real-valued snapshot image $m(x,y)$, it is necessary to encode each spin by a pre-specified real value. For the Ising model with $Q=2$, the up/down spins are most symmetrically represented by $\pm 1$. However, there is no symmetrical and natural way of mapping $Q=3$ different spins onto the real line. 

One may think that this leads to a potential problem, since different encodings choices will invariably lead to different snapshot spectra. Fortuntely, the most important quantity - the snapshot spectrum - turns out to be independent of this encoding. It has been shown that the encoding only affects the largest ensemble-averaged SVD eigenvalue\cite{matsueda2014comment,lee2014exact}, which has zero measure in the thermodynamic limit. Intuitively, the largest eigenvalue correspond to the ``background'' configuration with constant entries, and varies considerably according to the way the spins are represented. The other smaller eigenvalues contain information of the structures within the snapshot, and vary little with how the spins are represented. The interested reader may refer to Appendix \ref{app:encoding} for examples.

To further understand the nature of criticality, we shall hence focus in this paper the distribution of all snapshot eigenvalues \emph{except} the first. We shall call this truncated spectrum $\{\bar\Lambda_i\}$, $i=2,...,L$, normalized such that $\sum_{i=2}^L\bar\Lambda_i=1$. 
It is the distribution that contains information on critical behavior, and our random fractal ansatz will attempt to reproduce it. 

\section{The random fractal ansatz } 

In this section, we introduce the construction of our main object of interest, the random fractal-like ansatz image. Despite having only one tunable parameter, we shall see that its SVD spectrum and thus entropy can already closely resemble that of actual Ising/Potts snapshots. 

\begin{figure}
\includegraphics[scale=0.25]{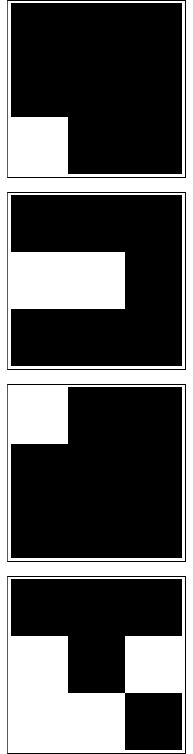}
\includegraphics[scale=0.58]{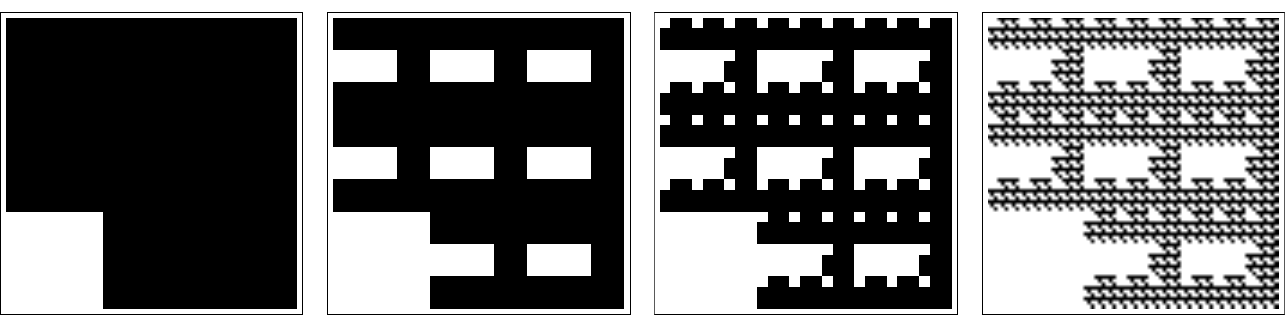}
\caption{(Color online) An illustration of the construction of the random fractal ansatz. The far left column depicts the four $3\times 3$ (i.e. $c=3$) unit cells $M_1,M_2,M_3,M_4$ from bottom to top. On the right is the sequence of tensor products of these unit cells in order of increasing detail: $M_4,M_3\otimes M_4,M_2\otimes M_3\otimes M_4$ and $M_1\otimes M_2\otimes M_3\otimes M_4$. We see that $M_4$ controls the overall shape of the fractal, while $M_3,M_2$ and $M_1$ controls its details at increasingly smaller scales. }
\label{fractexp}
\end{figure}

Since real systems at criticality exhibit self-similarity at different scales, we propose an ansatz image $\tilde M$ of the form
\begin{equation}
\tilde M=M_1\otimes M_2\otimes M_3 \otimes...\otimes M_l
\end{equation}
where each $M_j$, $j=1,...,l$, is a $c_1\times c_2$ pixel image (matrix) that we refer to as an unit cell. When $\tilde M$ is written out explicitly as a $L_1\times L_2$ matrix (defining $L_i=c_i^l$), $M_1$ controls its smallest scale features, while progressively larger scale features depend on $M_2$, $M_3$, etc. For the purpose of practical implementation, we impose a maximum iteration level $l$, with $M_l$ controlling details at the largest scale. This construction is illustrated in Fig. \ref{fractexp}.

Information on the structure of $\tilde M$ is contained in the spectrum of $\rho=\tilde M\tilde M^\dagger$, which is the squared SVD spectrum of the fractal ansatz image $\tilde M$. $\rho$ shall be termed as the reduced density matrix (RDM), in analogy to quantum entanglement where quantum states instead of rank-one images are ``entangled".

The next step is to specify the unit cells $M_j$ that make up $\tilde M$. We shall consider \emph{ensembles} of these unit cells, since individual snapshots do not respect the symmetries (i.e. rotations) of the critical physical system. To preserve scale invariance, each $M_j$ must be drawn from the same probability distribution. Specific ensembles of unit cells will be discussed in Sect. \ref{sect:decay}; here, we shall first attempt to understand how the choice of unit cell ensemble affects the important spectral properties of the RDM.

\subsection{Spectrum of the reduced density matrix (RDM)}
Most importantly, the eigenvalue distribution of $M_j$ must result in a RDM spectrum that resembles that of snapshots of real critical systems. For each $j$, there are $c=\text{min}(c_1,c_2)$ nonzero eigenvalues $m_j$ of $M_jM_j^\dagger$ (In this paper we will only include nonzero eigenvalues in the spectral function). Since the RDM $\rho=\tilde M\tilde M^\dagger=\otimes_{j=1}^l M_jM_j^\dagger$, the eigenvalues $\tilde m$ of $\rho$ are given by 
\begin{equation}
\tilde m= \prod_{j=1}^l m_j,
\end{equation}
Denote the spectral density function of $m_j$ as $p(m_j)$. Then $\tilde m$, which is distributed according to the RDM spectral function $\tilde p(\tilde m)$, is the product of $l$ independent, identically distributed random variables $m=m_j$, each distributed according to $p(m)$. Here and below, we shall always use $\tilde m$ for the eigenvalue of the whole RDM, and $m_j$ (or simply $m$ since the unit cells are scale-invariant) for the eigenvalue for an unit cell. 

Through the use of characteristic functions (see Appendix \ref{app:spectralCF}), the RDM spectral distribution $\tilde p(\tilde m)$ is related to that of the unit cell spectrum $p(m)$ via 
\begin{eqnarray}
\tilde p(\tilde m)
&\propto &\int_{-\infty}^\infty\tilde m^{-1-it}\left[\int_{0}^\infty m^{it}p(m)dm\right]^ldt
\label{convo2}
\end{eqnarray}
with the constant of proportionality most easily fixed by requiring that $Tr \rho=\sum \tilde m=L\int \tilde m p(\tilde m)d\tilde m=1$, $L=c^l$ where $c=\text{min}(c_1,c_2)$. 
Given unit cells $M_j$ drawn from known distributions, one can always compute the unit cell spectral function $p(m)$. From that, the spectrum of $\rho$ and hence snapshot entropy can be precisely determined via Eq. \ref{convo2}. A few examples of common $p(m)$ are given in Appendix \ref{app:spectral}.

\section{Limit of large number of iterations $l$} 

Although the random fractal ansatz described above can be defined for snapshots with any number of iterations $l$, of most physical interest is the large $l$ limit where $\tilde M$ looks manifestly scale-invariant. Fortuitously, elegant general results exist in this limit by virtue of the central limit theorem. We shall discover that a generic random fractal snapshot spectrum always exhibits an approximately scale-free regime and a disordered regime.

\subsection{Derivation of large $l$ RDM spectrum}
\label{RDMspec}

In the limit of large number of iterations $l$, the spectrum $\tilde p(\tilde m)$ of the RDM $\rho$ tends to a log-normal distribution \emph{regardless} of the choice of the unit cell. This general property is a direct consequence of the central limit theorem. Denote by $\sigma^2$ the variance of $x_j=\log m_j$, which is the logarithm of the unit cell eigenvalue:
\begin{equation} \sigma^2=\text{Var}[\log m_j]=\text{Var}[\log m]\end{equation}
With large $l$, the distribution of $\tilde x= \log \tilde m=\sum_j^l \log m_j$ tends towards a normal distribution with variance $l\sigma^2$. Hence the large $l$ RDM spectral distribution takes the form of a \emph{log-normal} distribution:
\begin{eqnarray}
\tilde P(\tilde m)=\lim_{l\gg 1}\tilde p(\tilde m)&=&\frac1{\tilde m \sigma \sqrt{2\pi l}}e^{-\left(\log(\tilde m L)+\frac{l\sigma^2}{2}\right)^2/2l\sigma^2}\notag\\
&=&\frac1{2\tilde m \sqrt{\pi\Sigma}}e^{-\left(\log(\tilde m L)+\Sigma\right)^2/4\Sigma},
\label{lognormal}
\end{eqnarray}
which is parametrized by the two independent parameters $\Sigma=\frac{l\sigma^2}{2}$ and $L=\text{min}(c_1^l,c_2^l)$, the effective system size. In the above, the normalization constraint $Tr \rho=1= L\int \tilde m P(\tilde m)d\tilde m$ has constrained the mean of the corresponding Gaussian distributed\footnote{The mean and variance of $\tilde m$ itself are $\frac1{L}$ and $\frac{e^{2\Sigma}-1}{L^2}$ respectively.} $\tilde x=\log\tilde m$ to be $-\log L -\Sigma$. 

Eq. \ref{lognormal} can be inverted to obtain explicit values for the eigenvalues $\tilde m=\tilde \lambda_1,\tilde \lambda_2,...,\tilde \lambda_L$, arranged in decreasing order. Note the tilde above $\tilde\lambda $ to avoid confusion with the unit cell eigenvalues. Since there are $i$ eigenvalues equal or larger than $\tilde\lambda_i$,  
\begin{equation}
\frac{j}{L}=\int_{\tilde\lambda_j}^{\infty}\tilde P(\tilde m)d\tilde m
\label{inv}
\end{equation}
which yields
\begin{equation}
\tilde\lambda_j=\frac1{L}\text{Exp}\left[2\text{Erfc}^{-1}\left(\frac{2j}{L}\right)\sqrt{\Sigma}-\Sigma\right]
\label{inv2}
\end{equation}
where $\text{Erfc}^{-1}(z)$ is the inverse of the complementary Error function $\text{Erfc}(z)=\frac{2}{\sqrt{\pi}}\int_x^\infty e^{-t^2}dt$, and $\Sigma =\frac1{2}l\sigma^2$. 

Physically, the variance $\sigma^2$ can be regarded as an intrinsic property of the shape of the $p(m)$ distribution, as well as a measure of the \emph{intrinsic disorder} in the ensemble of unit cells. This is because it is independent of the system size, which can change the normalization of $m$ but not the variance of $\log m$. The quantity $\Sigma=\frac1{2}l\sigma^2$, which scales as the number of iterations $l$, is thus a measure of the total ensemble disorder.


\subsection{Scale-invariant regime in the RDM spectrum}
\label{regime}

\begin{figure}
\includegraphics[scale=0.8]{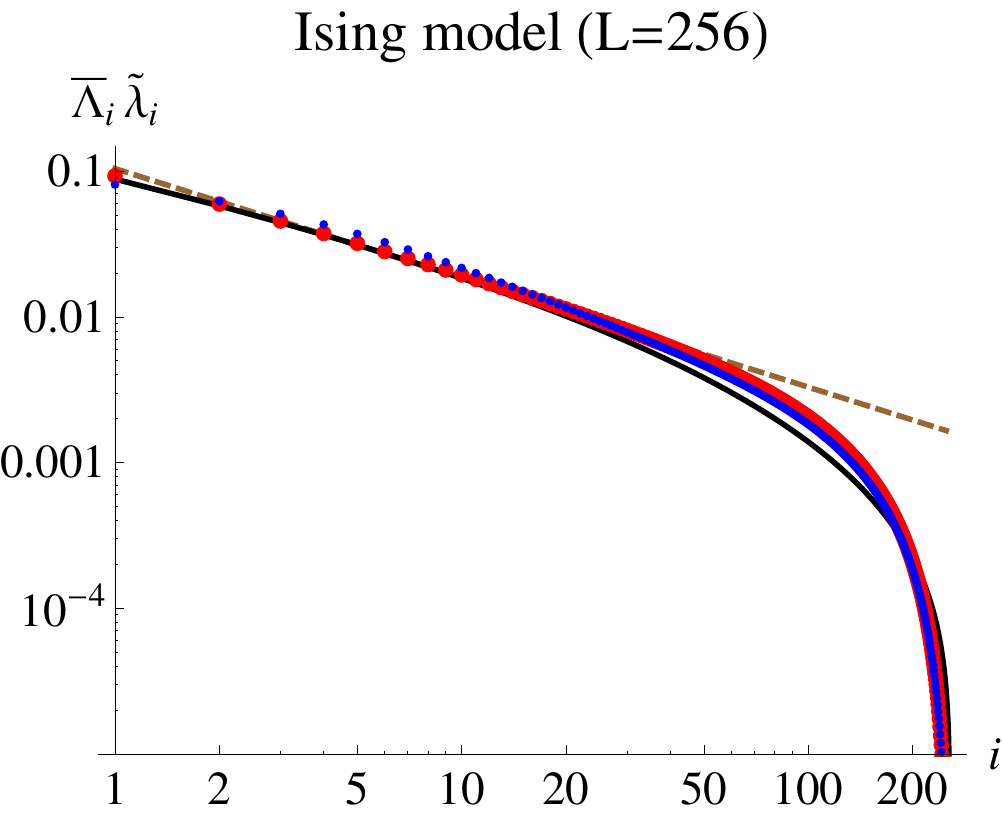}
\includegraphics[scale=0.94]{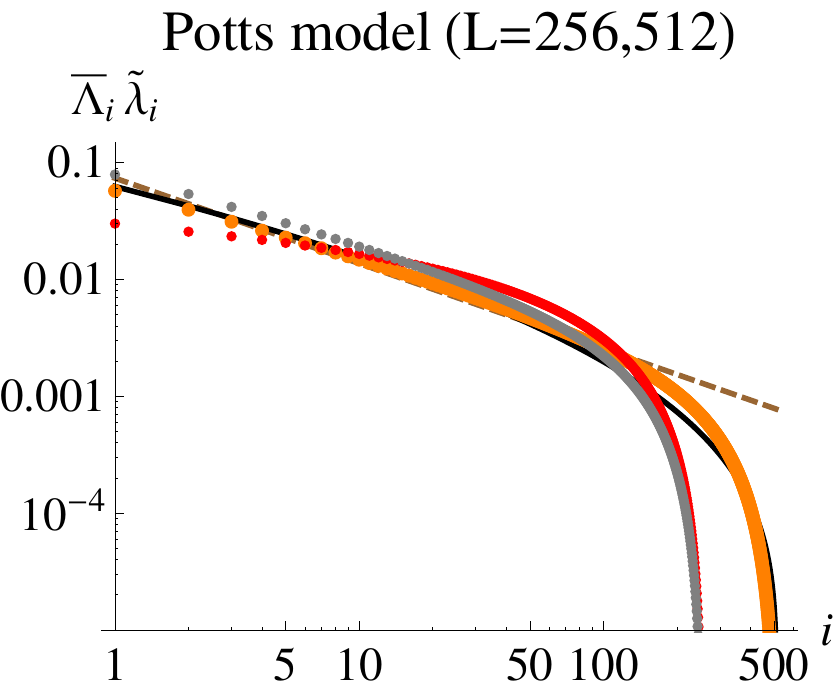}
\caption{(Color online) Log-log plots of the numerical snapshot spectra $\bar\Lambda_j$, $j=2,3,...,L$ of Ising and 3 states Potts models against the random fractal spectrum $\tilde \lambda_i$ with $\Sigma$ given by Eq. \ref{power4}. 
Above) The Ising model at $T=2.268\approx T_C$ (red) and $T=2.35>T_C$ (blue). The best agreement with the random fractal spectrum $\tilde\lambda_i$ (black curve) holds at the critical temperature $T_C$, where the error incurred is less than $0.1\%$. Below) The Potts model at two system sizes. Plotted for $L=256$ are the $T=1.93<T_C$ (red) and $T=1.99\approx T_C$ (gray) cases, while plotted for $L=512$ is the $T\approx T_C$ (orange) case. The agreement with the random fractal spectrum (solid black line) improves with system size, with an error of less than $0.1\%$. In both plots, the dashed brown line indicates the theoretically predicted power-law decay curve for larger $\bar\Lambda_i$, with exponents $\Delta = \frac{3}{4}$ and $\Delta=\frac{11}{15}\approx 0.733$ respectively. }
\label{lambda}
\end{figure}

Scale invariance is a hallmark of criticality, and should be characterized by a power-law decaying regime in the snapshot entropy. This is indeed observed in the snapshot spectra of our numerical simulations of critical Ising and 3 state Potts models, where the larger eigenvalues $\Lambda_i$ decay approximately as a power-law (see Fig. \ref{lambda}).

Following the arguments in Refs. \onlinecite{imura2014snapshot,matsueda2014comment}, this power law decay in the real snapshot spectra can be traced to the real-space power law decay of the RDM $[\rho]_{ik}=[\tilde M\tilde M^\dagger]_{ik}\sim |i-k|^{-\eta}$, where $\eta$ is the scaling exponent. Due to translational invariance, the RDM's eigenvalues $\tilde\lambda_j$ can be labeled by momenta $\frac{2 \pi j }{L}$. Hence the larger eigenvalues decay like\footnote{The integral in Eq. \ref{power1} can be evaluated via a Wick rotation into the ordinary Gamma integral.}:
\begin{eqnarray}
\tilde\lambda_j&\sim&\left|\sum_r \frac1{r^\eta}e^{ir\frac{2\pi j}{L}}dr\right|\notag\\
&\sim& \frac{\Gamma[1-\eta]}{\left(\frac{2\pi j}{L}\right)^{1-\eta}}\propto \frac1{j^{\Delta}} 
\label{power1}
\end{eqnarray}
where $\Delta =1-\eta$. Taking the continuum limit, this relation can also be inverted via Eq. \ref{inv} to yield 
\begin{equation}
\tilde p(\tilde m)\sim \tilde m^{-\alpha}
\label{power2}
\end{equation}
where $\alpha = 1+\frac1{\Delta }=\frac{2-\eta}{1-\eta}$. Due to the inherent disorder present in classical snapshots, we only expect this power law decay $\Lambda_i$ to hold asymptotically, i.e. Eqs. \ref{power1} and \ref{power2} should only hold for sufficiently large $\tilde m=\tilde \lambda_j$. This is clearly reflected in the linear regime of the log-linear spectral plots in Fig. \ref{lambda}. 

The random fractal ansatz also predicts an approximately power-law decaying regime for sufficiently large eigenvalues $\lambda_j$. From Eq. \ref{inv2}, the effective scaling exponent at the $j^{th}$ eigenvalue $\tilde m =\tilde \lambda_j$ is
\begin{eqnarray}
\Delta_j&=&-\frac{d\log \lambda_j}{d\log j}\notag\\
&=&\frac{2j}{L}\text{ Exp}\left[\text{Erfc}^{-1}\left(\frac{2j}{L}\right)^2\right]\sqrt{\pi\Sigma}
\label{power3}
\end{eqnarray}
which is approximately 
\begin{equation} 
\Delta \approx 0.6\sqrt{\Sigma}\approx 0.43\sigma \sqrt{l},
\label{power4}
\end{equation}
or $\Sigma\approx 2.71\Delta^2$, across the power-law regime at small $\frac{j}{L}$. Eqs. \ref{power3} and \ref{power4} presents a direct linear relationship between the critical exponent $\Delta$ and $\sigma$, the standard deviation of $\log m$ within the constituent fractal unit cells.

\subsection{Comparison with snapshot spectra of actual Potts models}

As evident from Fig. \ref{lambda}, there is excellent numerical agreement of the random fractal ansatz spectrum (Eq. \ref{inv2}) with that of an actual Ising/Potts snapshot, with $\Sigma=1.52, 1.46$ respectively as given by Eq. \ref{power4}. There are two salient regimes: the approximate scale-free regime at large $\bar\Lambda_j$ (small  $j$) with approximate power-law decay, and the disordered regime at small $\bar\Lambda_j$ (large $j$) with exponential decay $\tilde \lambda_j\sim \frac{e^{2\sqrt{\Sigma}-\Sigma}}{L}e^{-4\sqrt{\Sigma}\frac{j}{L}}$. This agreement is best in the neighborhood of the critical temperature $T_C$, where there is true scale invariance (up to limitations due to finite system sizes).

Previous works (Refs. \onlinecite{imura2014snapshot, matsueda2014comment}) accurately identified the scale-free decay exponent with $\Delta$ (brown dashed line in Fig. \ref{lambda}), but did not address its marked deviation from smaller eigenvalues. Physically, the smaller eigenvalues represent the ''noisy'' degrees of freedom arising from ensemble disorder, where scaling hierarchies become increasingly irrelevant. That this disordered regime is also accurately fitted by our random fractal ansatz ensemble is testimony to its correctness; one notes that there are various alternative random ensembles that do \emph{not} have the same qualitative spectra asymptotically, i.e. the random Wishart ensemble with Gaussian noise but without scale hierarchies, described in Eq. \ref{Wishartmi}. 

It is also worth noting that this good fit was achieved \emph{without} any fine-tuning of parameters, since $L$ and $\Delta\propto \sqrt{\Sigma}$ are already fixed by the system size and choice of Potts model.

\subsection{Snapshot Entropy}
We now turn towards various entropic measures of the random fractal spectrum, and analyze their agreement with snapshot data of actual Potts models. The entropy, which involves a sum over large numbers of SVD eigenvalues, allows more detailed study of the small-scale disordered regime where contributions from individual eigenvalues are extremely small.

\subsubsection{R\'{e}nyi entropy}
We first derive a general formula for the partially traced (i.e. finite scaling of) R\'{e}nyi entropy of the random fractal spectrum, from which all other entropic quantities can be derived. 
 
The R\'{e}nyi entanglement entropy traced over eigenvalues $\chi_i$ to $\chi_f$  is given by 
\begin{widetext}
\begin{eqnarray}
S_{\chi,n}&=&\frac{\log Tr_{\chi} \rho^n}{1-n}=\frac{\log \sum^{\chi_f}_{j=\chi_i} \tilde\lambda_j^n}{1-n}\notag\\
&\approx &\frac1{1-n}\log \left[L\int^{\tilde \lambda_{\chi_i}}_{\tilde\lambda_{\chi_f}}\tilde m^n\tilde P(\tilde m) d\tilde m\right ]\notag\\
&=&\log L -n\Sigma+\frac1{1-n}\log\left[ \frac1{2}\left(\text{Erf}\left[n\sqrt{\Sigma}-\text{Erfc}^{-1}\left(\frac{2\chi_f}{L}\right)\right]-\text{Erf}\left[n\sqrt{\Sigma}-\text{Erfc}^{-1}\left(\frac{2\chi_i}{L}\right)\right]\right)\right]
\label{renyi}
\end{eqnarray}
\end{widetext}
where $\text{Erf}(z)=1-\text{Erfc}(z)$. When the trace is taken over all the eigenvalues, one can also express the R\'{e}nyi entropy in terms of the characteristic function $\chi_{\tilde m}(t)$ (defined in Eq. \ref{convo}) via \begin{equation}
S_{L,n}=\frac{\log Tr \rho^n}{1-n}=\frac{\log L\langle \tilde m^n\rangle}{1-n} = \frac{\log L +\log \chi_{\tilde m}(-in)}{1-n}
\end{equation}

\subsubsection{Shannon entropy}
The Shannon entropy can be recovered from the R\'{e}nyi entropy by carefully considering the $n\rightarrow 1$ limit. 
We first consider its finite scaling (truncation) $S_\chi$, where only the first $\chi$ eigenvalues are traced over.
We have 
\begin{eqnarray}
S_\chi&=& -Tr_\chi \rho\log \rho\notag\\
&=&-\frac{d}{dn}\left(\log Tr_\chi \rho^n\right)|_{n=1}\notag\\
&=&-\frac{d}{dn}e^{(1-n)S_{\chi,n}}|_{n=1}\notag\\
&=&\sqrt{\frac{\Sigma}{\pi}}\left(e^{-\Omega^2}-e^{-E[\chi]^2}\right)\notag\\
&&+\frac{\Sigma-\log L}{2}\left(\text{Erf}[\Omega]-\text{Erf}[E[\chi]]\right)
\label{Schi}
\end{eqnarray}
where $E[i]=\sqrt{\Sigma}-\text{Erfc}^{-1}(2i/L)$ for $i\geq 1$, and $\Omega=\frac{\Sigma-\log L}{2\sqrt{\Sigma}}=\frac1{\sigma}\sqrt{\frac{l}{2}}\left(\frac{\sigma^2}{2}-\log c\right)$. Eq. \ref{Schi} was obtained by imposing in Eq. \ref{renyi} the cutoffs $\tilde \lambda_f=\chi$, and $\tilde \lambda_i=1$, which was necessary for avoiding spurious negative contributions to the entropy from the region $\tilde m=\tilde \lambda_i>1$. 


In the large $L$ limit, Eq. \ref{Schi} further simplifies to 
\begin{equation}
S_\chi|_{\log L>O(1)}\approx \frac{1+\text{Erf}[E[\chi]]}{2}(\log L -\Sigma)-\sqrt{\frac{\Sigma}{\pi}}e^{-E[\chi]^2}
\end{equation}

\begin{figure}
\includegraphics[scale=.93]{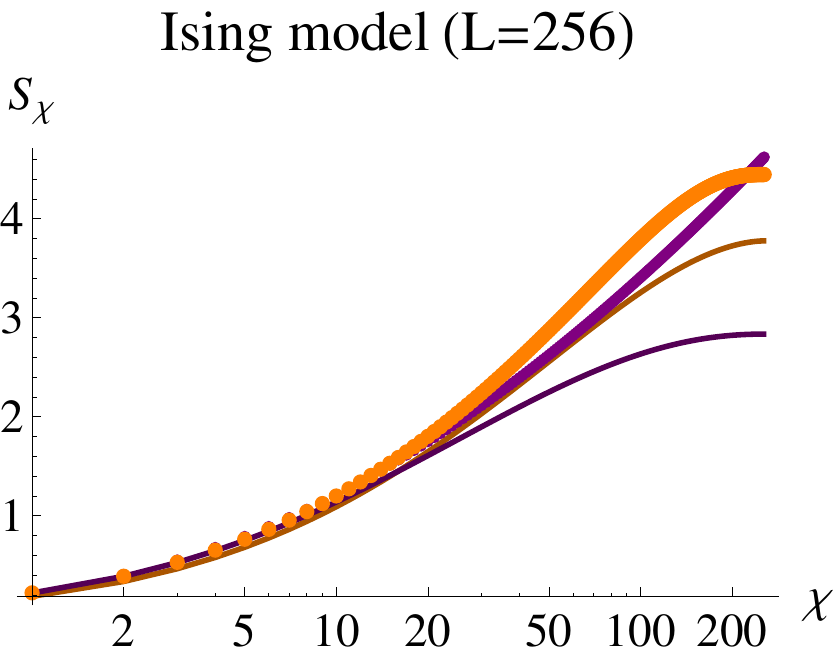}
\includegraphics[scale=.9]{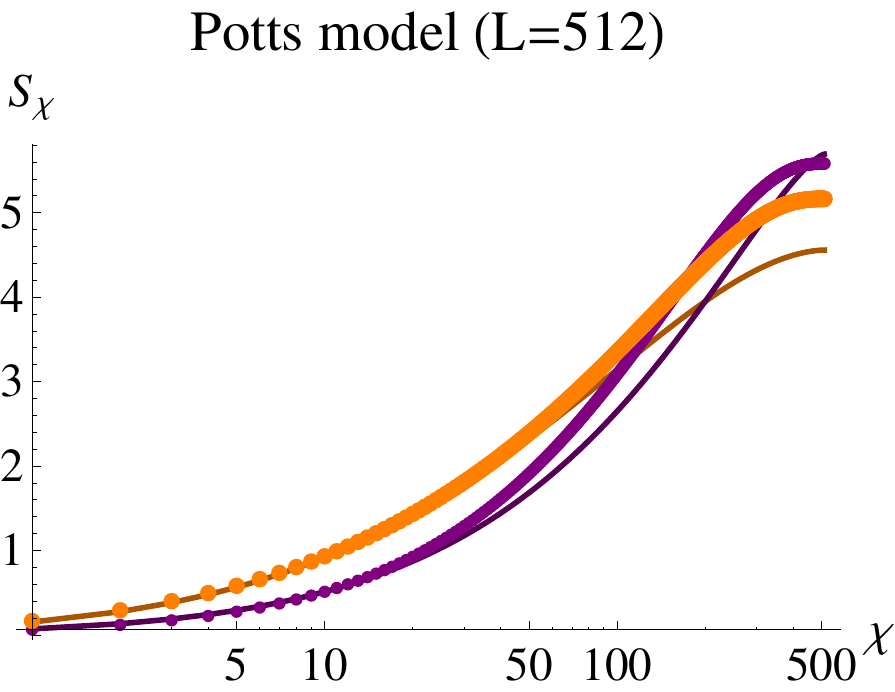}
\caption{(Color online) Log-linear plots of the finitely-truncated snapshot entropy $S_\chi$ for the Ising (Top) and $3$ states (Bottom) Potts models. The dark purple dots represent the subcritical cases $T=2,1.93<T_C$, and the orange dots represent the critical cases $T=2.268,1.99=T_C$ respectively. For both models, the agreement with the random ansatz entropy (Eqs. \ref{power4} and \ref{Schi}) is better at criticality (thin orange curve) than at below $T_C$ (thin purple curve). In general, the deviations are much larger at large $\chi$, where small deviations in each of the large number of eigenvalues add up.}
\label{Sfinite}
\end{figure}

As shown in Fig. \ref{Sfinite}, Eq. \ref{Schi} for the $S_\chi$ of random fractal snapshots agree fairly well with that of actual Ising/Potts snapshots, especially at small $\chi$ where the eigenvalues traced over mostly obey a power-law decay. As expected, the agreement is closest at the critical temperature, where the system is most accurately modeled by a random fractal. 

There is some discrepancy between the $S_{\chi}$ of the random fractal and that of actual Potts models at larger $\chi$. This is due to the large number of small eigenvalues that deviate slightly between the two: Although both systems exhibit quantitatively close eigenspectra as in Fig. \ref{lambda}, slight differences accumulate into a significant amount over $\sim L$ eigenvalues. Nevertheless, the random fractal ansatz still manages to reproduce the signature concavity of the $S_\chi$ curve at large $\chi$. 


Finally, one can recover the total snapshot entropy $S_L$ by setting\footnote{In this case $E(L)\rightarrow -\infty$, $\text{Erf}[E[L]]\rightarrow 1$} $\chi=L$ in $S_\chi$: 
\begin{eqnarray}
S_L&=& \sqrt{\frac{\Sigma}{\pi}}e^{\frac{-(\Sigma-\log L)^2}{4\Sigma}} +\text{Erfc}\left[\frac{\Sigma-\log L}{2\sqrt{\Sigma}}\right]\frac{\log L-\Sigma}{2}\notag\\
&=& \sqrt{\frac{\Sigma}{\pi}}\left(e^{-\Omega^2}+\sqrt{\pi}\Omega \;\text{Erfc}[\Omega]\right)
\label{SL}
\end{eqnarray}
which tends towards the simple expression 
\begin{equation}
S_L\rightarrow\log L-\Sigma = l\left(\log c -\frac{\sigma^2}{2}\right)
\end{equation}
in the limit of large $L$. 
In other words, the disorder in the random fractal reduces the amount of information by $2^{\frac{\sigma^2}{2}}$ bits per iteration.

There is some subtlety concerning whether $l$ or $c$ is taken as fixed when $L=c^l$ grows. If $l$ is regarded as fixed, so that the unit cell $c$ grows with $L$, $S_L\sim \log L-\frac{l}{2}\sigma^2$ with the coefficient of $\log L$ unity. This always has to be true for a distribution of the form $\tilde p (\tilde m)=Lf(L\tilde m)$, whose shape is independent of $L$. In this case we have $S_L=-L\int \tilde p(\tilde m)\tilde m\log \tilde m d\tilde m=\log L \int m' f(m')dm' -S_{L=1}=\log L  -\text{const.}$ 

If the unit cell size $c$ is instead regarded as fixed, so that $l$ grows like $\log L$, we simply have $S_L\propto l$, consistent with the expectation that the snapshot entropy should increase linearly with the number of independent iterations.


\subsubsection{Entropy production per iteration}
\begin{figure}
\includegraphics[scale=.9]{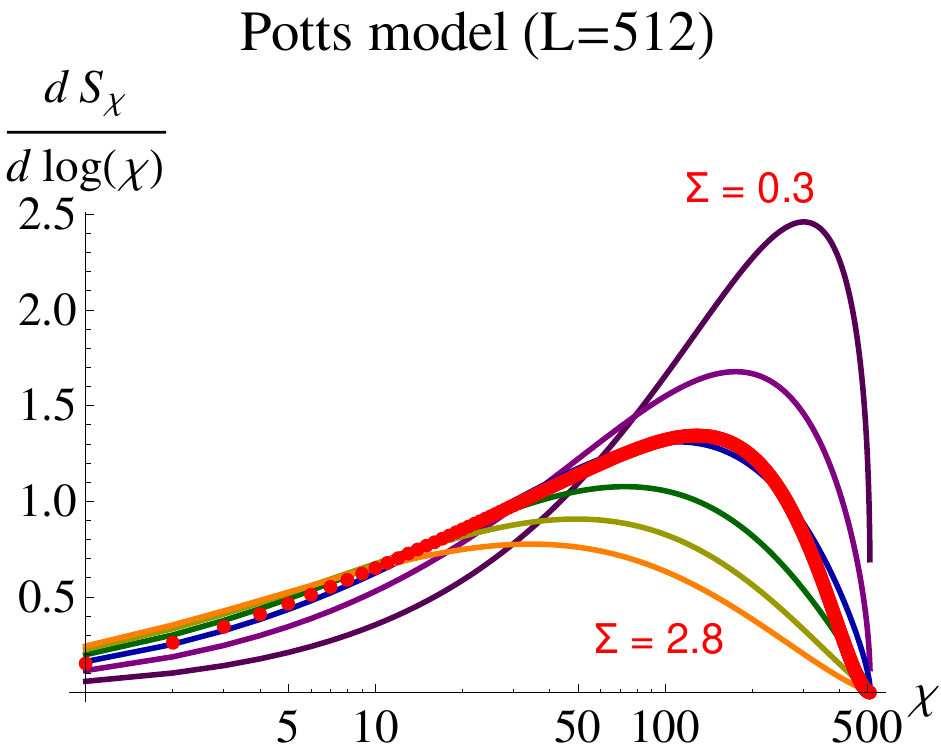}
\caption{(Color online) Plots of the snapshot entropy gained per iteration $\frac{d S_\chi}{d\log\chi}$ for $\Sigma=0.3,0.8,1.3,1.8,2.3,2.8$, arranged from top to bottom. There is more entropy, and hence effective rank, gained at smaller scales (large $\chi$) when $\Sigma$ is small. For comparison, the numerically obtained $\frac{d S_\chi}{d\log\chi}$ for the $3$ states Potts model is indicated by the red dots, where good agreement with the $\Sigma=1.3$ curve can be seen.  }
\label{SchiD}
\end{figure}

To understand the shape of the $S_\chi$ curve better, one can look at 
\begin{equation}
\frac{d S_\chi}{d\log\chi}=\frac{\chi}{L}e^{-\gamma}(\gamma+\log L),
\end{equation}
where $\log\chi$ is proportional to the number of iterations traced over, and $\gamma=\Sigma-2\sqrt{\Sigma}\text{ Erfc}^{-1}\frac{2\chi}{L}$. Physically $\frac{d S_\chi}{d\log\chi}$ is proportional to the snapshot entropy (and hence $\log[\text{rank}]$) gained per iteration. From Fig. \ref{SchiD}, we see that $\frac{d S_\chi}{d\log\chi}$ has strong peak at large $\chi$ for small $\Sigma$, but becomes more gently peaked at moderate $\log\chi$ for large $\Sigma$. Physically, snapshots of fractal ensembles with small $\Sigma$ show little randomness at large scales (small $\chi$), and thus accumulate less complexity (i.e. increase in rank) more slowly. Snapshots with large $\Sigma$ look more random at all scales, and hence look more complicated at larger scales. However, their randomness also serves to limit their effective complexity beyond a certain number of iterations, hence suppressing $\frac{d S_\chi}{d\log\chi}$ at small scales (large $\chi$). The fitting of $\frac{d S_\chi}{d\log\chi}$ with actual data from the critical Potts model also provides another avenue for determining $\Sigma$, which we show in Fig. \ref{SchiD} agree well with that from Eq. \ref{power4}. 

\section{Deviations from criticality }

\begin{figure}
\includegraphics[scale=.94]{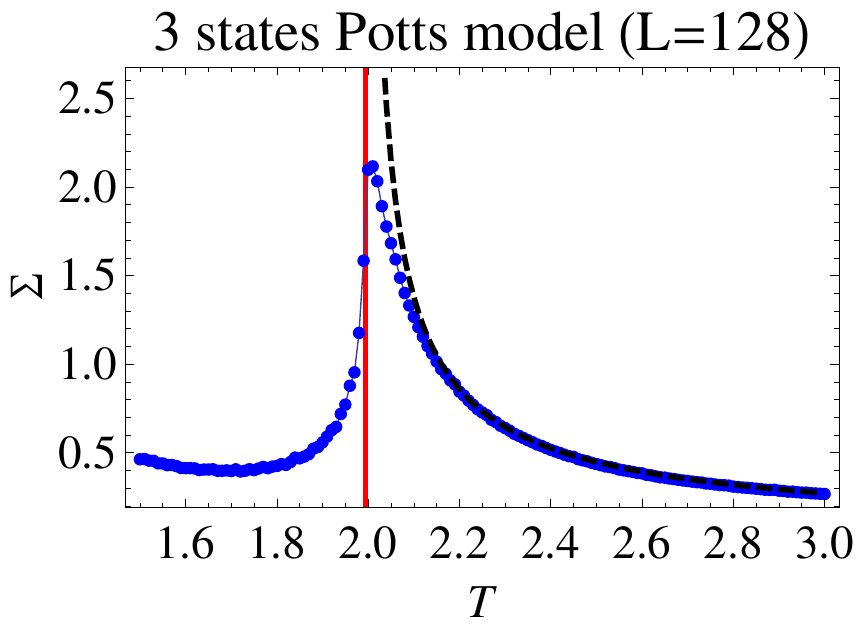}
\caption{(Color online) The variation of the best-fit $\Sigma$ (blue) as a function of temperature for the 3 states Potts model. $\Sigma$ rises dramatically around the critical temperature of $T_C=1.99$ (red line). Above $T_C$, the decay of $\Sigma$ can be fitted very accurately with a power-law curve of exponent $0.733=\Delta$ (black dashed line. We do not attempt to fit the variation below $T_C$ due to complications from the largest-eigenvalue condensation caused by large ``islands'', as elaborated in Appendix \ref{app:encoding}. }
\label{temp}
\end{figure}

Slightly away from the critical temperature, snapshot ensembles of the Ising/Potts models will still exhibit conformal invariance up to a certain scale. Hence we should still expect the random fractal ansatz to reproduce their snapshot spectra accurately up to a certain scale. Deviations from the criticality must be contained in $\Sigma$, the only free independent parameter of the ansatz (At criticality $\Sigma$ is fixed by $\Delta$). In Fig. \ref{temp}, we show how $\Sigma$ is ``renormalized'' away from the critical temperature for the 3 states Potts model, where its best-fit value (up to tiny errors due to finite size) exhibits a distinct power-law variation with the distance from criticality:
\begin{equation}
\Sigma \sim \frac{\Sigma_0}{(T-T_C)^\Delta}
\end{equation}
for $T>T_C$, where $T_C=\frac{2}{\log(1+\sqrt{3})}=1.99$. Notably, $\Delta=\frac{11}{15}=0.733$ is exactly the same exponent as that of the power-law decay in $\Lambda_i$. This agreement, which rests on the universality of critical phase transitions, provides another approach for detemining $\Delta$ from numerical snapshot data.

There is a sharp peak of $\Sigma$ around $T_C$, where criticality entails the longest correlation range. The lower values of $\Sigma$ away from criticality implies less randomness within the unit cells, which leads to ``clumpy''-looking snapshot with shorter-range order.

\section{Construction of unit cells}
\label{sect:decay}

We have just seen how the spectral properties, including various entanglement measures, depend on the unit cell ensemble $p(m)$. Of particular significance is the scaling exponent $\eta=1-\Delta$ of the snapshot spectrum, which is defined at the ensemble level. 
However, at the level of \emph{individual} snapshots, the correlation functions may possess a different scaling exponent $\eta'$ . 
In this section, we shall discuss how one can choose unit cells that produce the decay exponents $\eta'$ that agree with $\eta$ for each individual snapshot.




Recall that a random fractal ansatz image is given by
\[ \tilde M=M_1\otimes M_2\otimes...\otimes M_l \]
where each unit cell $M_p$ is a $c\times c$ matrix (assuming square unit cells for now). Let us denote the elements of $M_p$ by $M_p^a$, with $a$ labeling the $c^2$ elements within each $M_p$. We shall also refer to $M_1\otimes...\otimes M_p$ as a level-$p$ unit cell, since it is a $c^p\times c^p$ subset of $\tilde M$.

Consider two pixels (spins) $s_i$ and $s_j$ within the same level-$(k+1)$ unit cell, but in different level-$k$ unit cells. These two spins will be spatially separated by $\sim c^k$ sites (pixels). Explicitly, they can be written as $s_i=\prod_{p=1}^l M_p^{a_p}$ and $s_j=\prod_{p=1}^l M_p^{a'_p}$, with $k$ the lowest level such that $a_p=a_p'$ $\forall$ $p\geq k$. Denoting by $\langle ...\rangle_k$ the expectation value across $~c^k$ sites, the joint expectation (correlator) of $s_i$ and $s_j$ within a single snapshot is given by
\begin{widetext}
\begin{eqnarray}
\langle s_is_j\rangle_k&=&\frac1{c^{2(2(k-1)+(l-k))}}\frac1{2\binom {c^2}2}\sum_{\text{configs}}s_is_j\notag\\
&=&\left[\frac1{2\binom {c^2}2c^{2(l-k-2)}}\sum_{\text{level $k$ configs $a,a'$}}M_k^aM_k^{a'}\right]\prod_{p=1}^{k-1}\left(\sum_a M^a_p\right)^2\prod_{p=k+1}^{l}\left(\sum_a(M_p^a)^2\right)\notag\\
&=&\left[\frac{(\sum_a M_k^a)^2-\sum_a (M_k^a)^2}{c^2(c^2-1)}\right]\prod_{p=1}^{k-1}\left(\frac1{c^2}\sum_a M_p^a\right)^2\prod_{p=k+1}^{l}\left(\frac1{c^2}\sum_a(M_p^a)^2\right)\notag\\
&=&\left[\frac{c^4\langle\langle M_k\rangle \rangle^2-\sum_a (M_k^a)^2}{c^2(c^2-1)}\right]\prod_{p=1}^{k-1}\langle \langle M_p\rangle \rangle ^2\prod_{p=k+1}^{l}\left(\frac1{c^2}\sum_a(M_p^a)^2\right)
\label{correlator1}
\end{eqnarray}
\end{widetext}
where $\langle\langle M_p\rangle\rangle=\frac1{c^2}\sum_a M_p^a$ denotes an expectation \emph{within} a single level of unit cell. The above computation is broken up into three regimes: levels $p<k$, levels $p>k$ and level $k$. As previously specified, the matrix elements $M_p^a$ for the two spins must be equal above level $k$, different at level $k$, and are unrestricted below level $k$.

In the first line, we have summed over all the $c^2$ configuration at each fractal level $1,...,k-1$ for \emph{each} of the two spins, all the $c^2$ equal configurations at levels $k+1,...,l$ for both spins, and $2\binom {c^2}2=c^2(c^2-1)$ joint configurations at level $k$. 
In the second line, we split the configurations over the three regimes. 
In the third line, we expanded the expression for level $k$, where we subtracted off configurations $\sum_a (M_k^a)^2$ corresponding to the two spins being in the same $c^k$-sized cell. 

Next, we normalize\footnote{Note that the results of the previous sections are unaffected by this normalization, since they depend on $\sigma^2$, the variance of $\log \tilde m$ which is normalization-independent. } the unit cells via
\begin{equation}
\frac1{c^2}\sum_a (M_p^a)^2 =1 
\end{equation}
for all levels $p$. A normalization of squared elements avoids divergences from $M_p$ where the positive and negative elements sum to zero. With that,
\begin{equation}
\langle s_is_j\rangle_k=\left[\frac{c^2\langle\langle M_k\rangle \rangle^2-1}{c^2-1}\right]\prod_{p=0}^{k-1}\langle \langle M_p\rangle \rangle ^2
\label{correlator2}
\end{equation}
The above expression depend on the unit cell matrix elements exclusively through quantities of the form $\langle\langle  M_p\rangle\rangle^2$. Define $N_p=c^2\langle\langle  M_p\rangle\rangle^2$. Then
\begin{equation}
\frac{\langle s_is_j\rangle_{k+1}}{\langle s_is_j\rangle_k}=\left[\frac{c^2\langle\langle M_{k+1}\rangle \rangle^2-1}{c^2\langle\langle M_k\rangle \rangle^2-1}\right]\langle \langle M_k\rangle \rangle ^2=\frac1{c^2}\frac{N_{k+1}-1}{1-N_k^{-1}},
\label{correlator3}
\end{equation}
an expression expressing correlator ratios in terms of the average of the elements within an unit cell.

\subsection{Scale invariant case}

Recall that $\langle s_is_j\rangle_k$ is the correlator of two spins separated by $\sim c^k$ sites. In the scale invariant case, both sides of Eq. \ref{correlator3} should thus be independent of $k$. Denote it by a constant $r=c^2\frac{\langle s_is_j\rangle_{k+1}}{\langle s_is_j\rangle_k}$. The difference equation Eq. \ref{correlator3} can be systematically solved\footnote{This can be done via the auxiliary variable $w_k=\prod_j^k N_j$ and the indicial equation $w^2-(1+r)w+a=0$.} to yield
\begin{equation}
c^2\langle\langle M_k\rangle\rangle^2=N_k=\frac{(r-N_1)+(N_1-1)r^k}{(r-N_1)+(N_1-1)r^{k-1}}.
\label{Nk}
\end{equation}
There exists a fixed point $N_k=r$ for all $k$, corresponding to a fractal image where the average of the unit cell elements $\langle\langle M_k\rangle\rangle$ takes on the same value $\sqrt{\frac{\langle s_is_j\rangle_{k+1}}{\langle s_is_j\rangle_k}}$ at all levels $k$.

Apart from extremely fast correlator decays with $\frac{\langle s_is_j\rangle_{k+1}}{\langle s_is_j\rangle_k}<\frac1{c^2}$, $r$ shall be greater than unity and $N_k$ converges to the fixed point $r$ as $k$ increases. In other words, we have
\begin{equation}
\sum_a M_k^a = c\sqrt{N_k}\rightarrow c^2\sqrt{\frac{\langle s_is_j\rangle_{k+1}}{\langle s_is_j\rangle_k}}
\label{correlator4}
\end{equation}
as $k$ increases, with departures from Eq. \ref{correlator4} fitting the ``boundary condition'' at $N_{k=1}$.  

\subsubsection{Power-law decaying correlators}
\label{powercorr} 

\begin{figure}
\includegraphics[scale=0.46]{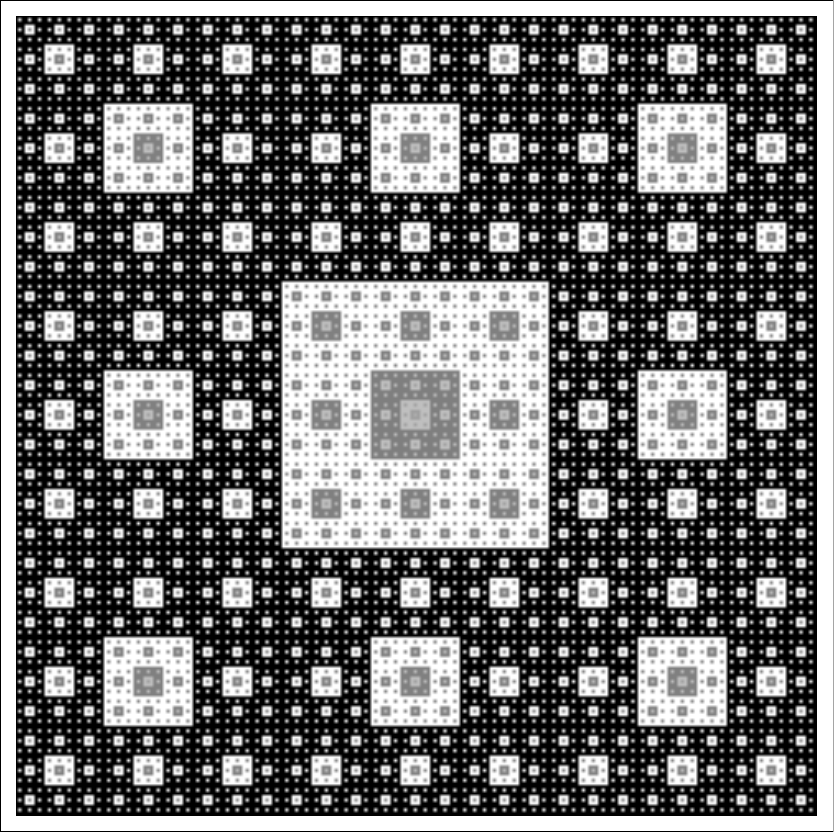}
\includegraphics[scale=0.54]{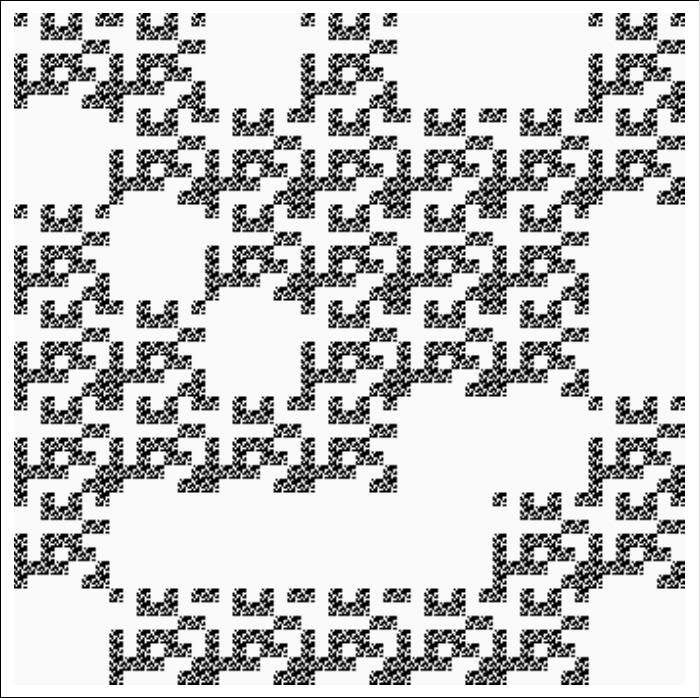}
\caption{(Color Online) Left) An $l=5$-iteration fractal image with unit cells of the form Eq. \ref{3by3}. The central pixels $\epsilon_k$ of each iteration are chosen such that the spatial correlators $\langle s_is_j\rangle_k\propto x^{-1/4}$, i.e. decay with a scaling exponent of $\eta'=\frac1{4}$, the exponent of the Ising model. Explicitly, they are $\epsilon_1=1,\epsilon_2=-0.4, \epsilon_3=-0.48$ and $\epsilon_4=-0.49,\epsilon_5=-0.5$. Right) An $l=3$-iteration random fractal image with $7\times 7$ unit cells having $30$ ones and $19$ zeroes each. This ratio of ones/zeroes leads to a scaling exponent of $\eta'\approx 0.252\approx \frac1{4}$. Although this particular random fractal is not scale and rotationally invariant, ensembles of it should be.
}
\label{corr}
\end{figure}

Consider a power law decay 
\begin{equation}
\langle s_is_j\rangle_k\propto x^{-\eta'}=c^{-k\eta'}
\end{equation}
where $x=c^k$ is the typical distance between the two spins at scale $k$, and $\eta'$ the scaling exponent. Substituting this in Eq. \ref{correlator4}, we obtain
\begin{equation} 
\sum_a M_k^a = c^{2-\eta'/2},\;\;\;\;\;\sum_a (M_k^a)^2=c^2,
\end{equation}
i.e.
\begin{equation}
\eta'=2+\log_c\frac{\sum_a(M_k^a)^2}{\left(\sum_a M_k^a\right)^2}
\label{correlator5}
\end{equation}
As a first illustration, we specialize to the case of $3\times 3$ unit cells producing a Sierspinski carpet: \begin{equation}
M_k=\frac{3}{\sqrt{8+\epsilon^2_k}} \left(\begin{matrix}
 & 1 & 1 & 1 \\
 & 1 & \epsilon_k & 1 \\
 & 1 & 1 & 1 \\
\end{matrix}\right).
\label{3by3}
\end{equation}
Eq. \ref{correlator5} requires that $\frac{8+\epsilon_k}{\sqrt{8+\epsilon^2_k}}=3^{1-\frac{\eta'}{2}}$. An example is shown in Fig. \ref{corr}.

For the second illustration, consider unit cells with only $\pm$ entries, representative of up/down spins. Eq. \ref{correlator5} then tells us that $\eta'=2\log_c\left(\frac{U+D}{U-D}\right)$, where $U,D$ are respectively the number of up/down spins. If the up/down spins are instead encoded by matrix entries $1$ and $0$ repsectively, we will have $\eta'=2-\log_c U$. This is also illustrated in Fig. \ref{corr}.

\subsection{Relation between the scaling exponents at the single snapshot and ensemble levels}

At the level of a single snapshot, Eq. \ref{correlator5} expresses the correlator decay exponent $\eta'$ in terms of the unit cell matrix elements:
\begin{equation}
\eta'=2+\log_c\frac{\sum_a(M_k^a)^2}{\left(\sum_a M_k^a\right)^2}
\label{correlator6}
\end{equation}
For an explicitly scale-invariant snapshot, the $M_k^a$'s and hence $\eta'$ are automatically constant for all scale levels $k$. 

To restore rotational invariance and continuous scale invariance, we will need to consider an ensemble of $M_k$. This ensemble must respect Eq. \ref{power4} and Eq. \ref{correlator6}, i.e. also possess an eigenvalue spectrum $p(m)$ such that 
\begin{equation}
\eta'=\eta = 1- \Delta\approx 1- 0.43 \sigma \sqrt{l}
\label{correlator7}
\end{equation}
where $\sigma^2$ is the variance of $\log m$ (See Sect. \ref{RDMspec}). Such an ensemble can be constructed by first listing down all the matrices of a certain type satisfying Eq. \ref{correlator6}, such as those described in Sec. \ref{powercorr}. Next, these matrices can be ascribed ensemble probabilities based on their eigenvalue spectra, such that the ensemble averaged spectral distribution obeys Eq. \ref{correlator7}.

Note that for generic ensembles, scale invariance of the correlator does not automatically imply scale invariance of the RDM spectrum at the ensemble level. In fact, the spectra of individual scale invariant fractals possess a series of step-like plateaus (see Ref. \onlinecite{lee2014exact} for illustrations), and do not automatically average out into a power-law spectrum unless our random fractal ansatz, for instance, is used.



\section{Conclusion}

Motivated by the numerical ease of computing classical configurations and inspired by holographic duality, we studied the entanglement properties of the classical configurations (snapshots) of Ising and 3 states Potts models through their SVD spectra. We went beyond previous numerical investigations by introducing a rigorously defined ensemble of fractal images whose spectra agree very well with those from actual Potts snapshots, with \emph{no tunable parameters} at criticality\footnote{$L$ and $\Sigma$ are fixed by the system size and decay exponent respectively.}. Mathematically, these images are iterated Kronecker products of ``unit cells'' random drawn from an ensemble with a pre-defined spectra distribution $p(m)$. 

In the thermodynamic limit of a large number of iterations $l$, the Central Limit Theorem dictates that the fractal image spectrum tends to the expression given by Eq. \ref{inv2}, dependent only on the system size and $\Sigma=\frac1{2}l\sigma^2$, where $\sigma^2=\text{Var}(\log m)$. The universality of Eq. \ref{inv2} is key to the robustness of the ansatz in fitting actual snapshot data, which consists of effective unit cells with undetermined spectra. From Eqs. \ref{power3} and \ref{power4}, one can also extract the approximate scaling exponent from $\Sigma$, which is a property of the microscopic degrees of freedom within the unit cells. Furthermore, Eq. \ref{inv2} also accurately extrapolates down to very small scales where disorder dilutes scale invariance but restores rotation invariance, a regime ignored by previous works.

To further understand and quantify the complexity within the snapshot ensemble, we systematically derived analytic expressions for various snapshot entropies. By looking at the entropy production per iteration, for instance, we discovered that while snapshot ensembles with more heterogeneous unit cells contain more complexity (entropy) at large scales, their greater disorder also limits the entropy contributions below a certain scale. 

In all, there are three ways of extracting the critical exponent $\Delta$ of the system: 1) Directly from the decay exponent of the snapshot spectrum, 2) from the entropy production per iteration and 3), the decay exponent of $\Sigma$ above the critical temperature. Once $\Delta$ is determined, one can also explicitly construct unit cells obeying Eq. \ref{correlator4} such that the spatial correlation within each fractal image decays like $\eta'=1-\Delta$.

\acknowledgements
CHL thanks Mark H. Jhon and Bo Yang for stimulating discussions. This work was supported by JSPJ Kakenhi Grant No.15K05222 and No.15H03652.

\appendix

\section{Spectral properties of the reduced density matrix}
\subsection{Derivation of $\tilde p(\tilde m)$ of the RDM}
\label{app:spectralCF}

To find $\tilde p(\tilde m)$ of the RDM, note that the characteristic function of the sum of independent random variables is equal to the product of their individual characteristic functions. Specifically, define
\[ x_j= \log m_j, \;\;\;\;\;   \tilde x=\log \tilde m, \]
with spectral distribution functions $q(x_j)=e^{x_j}p(e^{x_j})$ and $\tilde q(\tilde x)=e^{\tilde x}\tilde p(e^{\tilde x})$. Then $\tilde x=\sum_{j=1}^l x_j$ has a characteristic function $\chi_x(t)=\mathbb{E}(e^{itx})=\int_{-\infty}^\infty e^{itx}q(x)dx $ obeying 
\begin{equation}
\chi_{\tilde x}(t)=\prod_j \chi_{x_j}(t)=[\chi_x(t)]^l
\label{convo}
\end{equation}
since the $x_j$'s are all identically distributed\footnote{When $l=2$, the above just reduces to the usual convolution formula for Fourier transforms.}. The spectral distribution of the RDM $\rho$ can thus be computed from the inverse Fourier transform of $\chi_{\tilde x}(t)$:
\begin{eqnarray}
\tilde p(\tilde m)&=&\frac{\tilde q(\log \tilde m)}{\tilde m }\notag\\
&\propto&\int_{-\infty}^\infty e^{-(1+it)(\log \tilde m) }[\chi_{ x}(t)]^ldt\notag\\
&=&\int_{-\infty}^\infty\tilde m^{-1-it}\left[\int_{0}^\infty m^{it}p(m)dm\right]^ldt
\label{convo2a}
\end{eqnarray}

\subsection{Examples of distributions}
\label{app:spectral}

\subsubsection{Power-law distribution}
Unit cells with power-law distributed spectrum $p(m)=(\alpha-1) m^{-\alpha}$, $\alpha>1$ result in a log-Gamma-distributed RDM spectrum $\tilde p(\tilde m)=\frac{(\alpha-1)^l}{(l-1)!}\tilde m^{-\alpha}(\log \tilde m)^{l-1}$, where $l$ is the number of tensor product iterations. This can be directly derived from Eq. \ref{convo2a}.

Note that the effect of the iterations is to introduce products with $\log \tilde m$, which serve to further enhance contributions with very small $\tilde m$.

\subsubsection{Gaussian distribution}
Another common example is that of large $c_1\times c_2$ unit cells $M$ with Gaussian distributed elements of variance $s^2$. Then $\frac1{c_2}MM^\dagger$ is of Wishart-type 
with spectral function\cite{imura2014snapshot,matsueda2014comment}
\begin{eqnarray}
p(m)=\frac{1}{2\pi s^2}\sqrt{\left(\frac{m_+}{m}-1\right)\left(1-\frac{m_-}{m}\right)},
\label{pmgaussian}
\end{eqnarray}
$m_\pm=s^2\left(1\pm\sqrt{\frac{c_1}{c_2}}\right)^2$, and characteristic function 
\begin{equation}
\chi_x(t)=\frac{c_2}{\left(\sqrt{c_1}+\sqrt{c_2}\right)^2}{}^2 F_1 \left(\frac{3}{2},1-4its^2,3,\frac{4\sqrt{c_1c_2}}{\left(\sqrt{c_1}+\sqrt{c_2}\right)^2}\right)
\end{equation}
where ${}^2 F_1(a,b,c,z)$ is the ordinary Hypergeometric function. A closed form solution still exists for the characteristic function $\chi_x(t)$ of $x=\log m$, but the final RDM spectrum $\tilde p(\tilde m)$ can only be determined through numerically integration of $\chi_x(t)$.

The ordinary (or Gaussian) Hypergeometric function ${}^2 F_1(a,b,c,z)$ are defined by 
\begin{equation}
{}^2 F_1(a,b,c,z)=\sum_{n=0}^\infty \frac{\Gamma(a+n)\Gamma(b+n)\Gamma(c)}{\Gamma(a)\Gamma(b)\Gamma(c+n)}z^n
\end{equation}
for $|z|<1$, with $\Gamma(n)=\int_0^\infty x^{n-1}e^{-x}dx$. 

Further simplification is possible in the case with $s=\frac1{2}$, so that $a=\frac{3}{2}$,$b=1-it$ and $c=3$. In this case
\begin{equation}
\chi_x(t)=\frac1{4}\sqrt{\frac{c_2}{c_1}}\sum_{n=0}^\infty \binom {2n+2}{n+1}\binom {n-it}{n}\frac{z^{n+1}}{4^n(n+2)}
\end{equation}
where $z=\frac{4\sqrt{c_1c_2}}{(\sqrt{c_1}+\sqrt{c_2})^2}$. The symmetric case where $c=c_1=c_2$ corresponds to $z=1$, which is just outside the circle of convergence of the above series. However, it can be shown from direct integration of the definition of $\chi_x(t)$ that 
\begin{equation}
\chi_x(t)|_{c_1=c_2}=\frac1{\sqrt{\pi}}\frac{\Gamma\left(\frac1{2}+it\right)}{\Gamma(2+it)}
\end{equation}

Going back to the general case, it can be shown by inverting Eq. \ref{pmgaussian} that Wishart eigenvalues $\tilde m_i$ satisfy
\begin{equation}
\frac{i}{L}=\frac{2}{\pi}\left(\cos^{-1}\frac{\sqrt{\tilde m_i}}{2\sigma}-\frac{\sqrt{\tilde m_i}}{2\sigma}\sqrt{1-\frac{\tilde m_i}{4\sigma^2}}\right)
\label{Wishartmi}
\end{equation}
for a $L\times L$ system, which is fundamentally different from the relation obyed by our random fractal ansatz (Eq. \ref{inv2}).

\subsubsection{Log-normal distribution}
Units cells with log-normal distributed $p(m)$ (defined in Eq. \ref{lognormal}) will yield an RDM spectrum that is still log-normal. This is because products of log-normal distributed variables are still log-normal distributed, just as sums of Gaussian distributed variables are still Gaussian distributed. The logarithm of a Cauchy or L\'{e}vy distributed variables also possess similar invariance properties.

If $\log m$ has a mean of $\mu$ and variance of $\sigma^2$, $\log\tilde m$ will have a mean of $l\mu$ and variance $l\sigma^2$.

\section{$\sigma^2$ for common distributions}
\label{app:sigma}
From the main text, the intrinsic disorder in the unit cell spectrum can be quantified by $\sigma^2$. For some common distributions, the value of $\sigma^2$ are 
\begin{itemize}
\item $\sigma^2=\frac1{(\alpha-1)^2}$ for a scale-free distribution with $p(m)\propto m^{-\alpha}$ for $m\geq m_{min}$, and $p(m)=0$ otherwise. This result holds for all $\alpha >1$, but is extremely sensitive towards the exact shape of the small $m$ cutoff when $1<\alpha<2$.

\begin{figure}
\includegraphics[scale=.85]{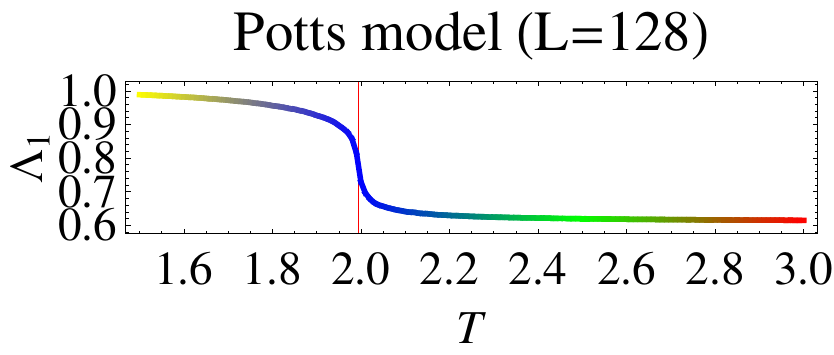}
\includegraphics[scale=.85]{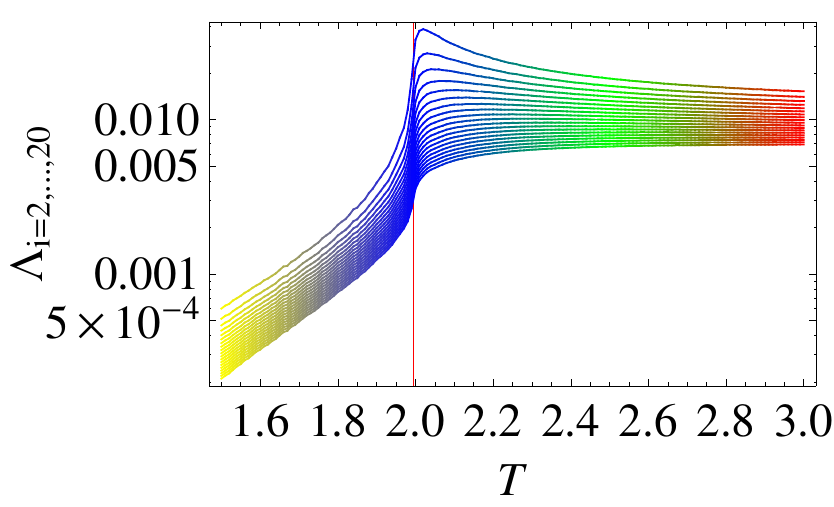}
\caption{(Color online) Top) The largest ensemble averaged snapshot eigenvalue $\Lambda_1$ for the 3 states Potts model. Bottom) The next $19$ ensemble averaged snapshot eigenvalue $\Lambda_2,...,\Lambda_{20}$ plotted on a log scale. They are all separated from $\Lambda_1$, and are exponentially suppressed for $T<T_C=1.99$. This suppression is due to the spatial correlation length being proportional to $T_C-T$. }
\label{Potts-1-20}
\end{figure}

For instance, consider the infinitely iterated Log-normal distribution (Eq. \ref{lognormal}) with $\Sigma\propto l\rightarrow \infty$ as $l\rightarrow \infty$. In the main text, it is shown to tend towards an ($l=1$ iteration) scale-free distribution with $\alpha=\frac{3}{2}$, but with a slightly smoother cutoff. Now, this scale-free distribution has $\Sigma=\frac1{2}\sigma^2=\frac1{2}\left(\frac{3}{2}-1\right)^{-2}=2$, which is finite, while the infinitely-iterated Log-normal distribution which approximates it, has infinite $\Sigma$.
\item  $\sigma^2=1+\frac{\pi^2}{3}\approx 4.29$ for a large $c\times c$ (square) unit cell with Gaussian distributed elements. 
\end{itemize}

\section{The largest SVD eigenvalue}
\label{app:encoding}

As previously discussed in Refs. \onlinecite{matsueda2012holographic,matsueda2015proper}, the largest SVD eigenvalue for the snapshots depend on how the various spins are encoded, and become especially prominent below $T_C$ where the spins form large islands. For illustration, the eigenvalues $\Gamma_1$ and $\Gamma_2,...,\Gamma_{128}$ of the 3 states Potts model are displayed in Fig. \ref{Potts-1-20} as a function of temperature. The particular encoding used was values $0,1$ and $2$ representing the three types of spins. 

The largest eigenvalue $\Lambda_1$ is totally separate from the other eigenvalues which are exponentially small. While it can be used as a clear indicator of the phase transition temperature $T_C$, it obviously does not obey the continuous decay curve of random fractal eigenvalues. Hence for numerical comparison with the random fractal ansatz, it should be excluded from the other eigenvalues, which are then separately normalized as $\bar \Lambda_i$.




\bibliography{ehm,entanglement}

\end{document}